\documentstyle[12pt]{article}
\newcommand{\BEQ}{\begin{equation}}
\newcommand{\NEQ}{\end{equation}}
\newcommand{\nn}{\nonumber}

\def\al{\alpha }

\def\Iuu{{\bf I}\,}
\def\Juu{{\bf J}\,}

\def\Omuu{{\bf \Omega}\, }
\def\Siuu{{\bf \Sigma}\, }

\def\ebf{{\bf{e}}}
\def\ybf{{\bf{y}}}
\def\ybt{\tilde{\bf{y}}}
\def\xv{\vec{x}}

\def\bth{{\hat{\beta}}}

\def\al{{\alpha}}

\def\th{{\theta}}

\def\btvh{\hat{\vec{ \beta}}}
\def\btv{\vec{ \beta}}

\def\xv{\vec{x}}

\def\np{\vfill\eject}       
\def\ni{\noindent}

\def\Xuu{{\bf X}\,}
\def\Siuu{{\bf \Sigma}\, }

\def\Wuu{{\bf W}}

\def\btv{\vec \beta}

\def\Cuu{{\bf C}\,}
\def\Duu{{\bf D}\,}
\def\Euu{{\bf E}\,}
\def\Fuu{{\bf F}\,}
\def\Guu{{\bf G}\,}
\def\Iuu{{\bf I}\,}
\def\Kuu{{\bf K}\,}
\def\Luu{{\bf L}\,}
\def\Muu{{\bf M}\,}
\def\Puu{{\bf P}\,}

\def\Xuu{{\bf X}\,}
\def\Zuu{{\bf Z}\,}
\def\Xbf{{\bf X}\,}
\def\Zbf{{\bf Z}\,}

\def\Dl{{\bf \Delta}}

\def\albf{\mathop{\mbox{\boldmath $\alpha$}}}
\def\albh{\hat{\mathop{\mbox{{\boldmath $\alpha$}}}}}
\def\bth{\hat{\mathop{\mbox{{\boldmath $\beta$}}}}}
\def\btbf{{\mathop{\mbox{\boldmath $\beta$}}}}
\def\Ombf{\mathop{\mbox{\boldmath $\Omega$}}}
\def\Sibf{\mathop{\mbox{\boldmath $\Sigma$}}}

\def\Sik{\mathop{\mbox{{\boldmath $\Sigma$}$_k$}}}

\def\Sinvk{\mathop{\mbox{{\boldmath $\Sigma$}$^{-1}_k$}}}
\def\Sinvl{\mathop{\mbox{{\boldmath $\Sigma$}$^{-1}_{\ell}$}}}

\begin{document}
\bibliographystyle{plain}

\begin{center}
REDUCTION OF RESTRICTED \\ 
MAXIMUM LIKELIHOOD FOR\\
RANDOM COEFFICIENT MODELS \\
K.S.~Riedel \\
Courant Institute of Mathematical Sciences \\
New York University \\
New York, N.Y.~10012-1185
\end{center}
\begin{abstract}
The restricted maximum likelihood (REML) estimator
of the dispersion matrix for random coefficient models 
is rewritten in terms of
the sufficient statistics of the individual regressions.

\ni

\np
\end{abstract}
{\bf I.  Reduction of the Restricted Maximum Likelihood}

Restricted maximum likelihood (REML) estimators are widely used
to estimate the free parameters in the dispersion matrix
for mixed models in structured linear regressions
$[1,4-10,13,19]$. The REML
estimator is the maximum likelihood estimate of the parameters
which uses only the information not contained in the estimate
of the the regression vector, and thereby automatically corrects
for the degrees of freedom which are lost in estimating the regression
vector. 

We consider the subclass of mixed models where the observations 
are grouped by individual/category, and the observations
are uncorrelated across individuals. The resulting covariance matrix has a 
block diagonal structure. Random coefficient (RC) models are a popular
subclass of mixed models where a subset of the regression coefficients
varies randomly across individuals $[2,3,9,12-21]$. 
In this note, we simplify the REML estimator
of the random coefficient model 
using linear algebra identities.  By expressing the likelihood in terms
of the sufficient statistics of the individual regressions, 
the REML degree of freedom correction may be better understood.

A mixed linear model consists of a fixed effects vector, $\alpha$,
a fixed effects covariate matrix, $\Xbf$, 
a random effects covariate matrix, $\Zbf$, and three random vectors;
the measurement vector, $y$, the measurement error vector, $\ebf$,
and the random effects vector, $\btbf$, which satisfy the linear equation:
$\ybf = {\bf X} \albf + {\bf Z} \btbf + \ebf $. We restrict our consideration
to linear mixed models with a block structure: $\ybf$, $\btbf$ and $\ebf$ 
consist of $N$ statistically independent subvectors,
$\ybf^T = (\ybf_1^T,\ldots \ybf_N^T)$, 
$\btbf^T = (\btbf_1^T,\ldots \btbf_N^T)$ 
and $\ebf^T = (\ebf_1^T,\ldots \ebf_N^T)$. 
We also require $\ybf$, $\btbf$ and $\ebf$ 
to be Gaussian random variables. Thus the $k$th individual,
$\ybf_k$, is a Gaussian random variable with the block linear mixed model
structure:  
\BEQ \label{E1}
\ybf_k = {\bf X}_k \albf + {\bf Z}_k {\btbf}_k + \ebf_k \ , 
\NEQ 
where ${\bf X}_k$ and ${\bf Z}_k$ are covariate matrices of\def\albh{\hat{\mathop{\mbox{{\boldmath $\alpha$}}}}}
dimension $n_k \times p$ and $n_k \times q$ respectively. 
We allow both ${\bf X}_k^t {\bf X}_k$ and ${\bf Z}_k^t {\bf Z}_k$ to be
singular, and denote their respective ranks by $p_k$ and $q_k$. 
$\ebf_k$ is a normally distributed  random $n_k$-vector of measurement errors 
with zero mean and covariance ${\bf E}[\ebf_k \ebf_k^t] =
\sigma_k^2 \Iuu_{n_k}$, with $\sigma_k^2 > 0$. 
$\albf$ is the $p$-vector of fixed effects, and
$\btbf_k$ is the $q$-vector of random effects. We assume that
$\btbf_k$ is 
a zero mean Gaussian random
vector with a $q \times q$ covariance matrix, $\Duu( \theta)$, 
which may be singular. $\theta$ is an unknown vector that
parameterizes $\Duu( \theta )$, and $\theta$ is an element in a known,
compact, parameter region, $\Theta$. 
Thus the covariance of the $k$th individual
satisfies
\BEQ \label{E2}    
\Sik = \sigma_k^2 \Iuu_{n_k} + {\bf Z}_k {\bf D}(\theta)\Zbf_k^t 
\ . \NEQ 
We assume that both $\ebf_k$ and $\btbf_k$ are independent
between individuals.
We require that $\displaystyle{\sum_{k=1}^N} {\bf X}_k^t \Sinvk{\bf X}_k$ 
has full rank. We also assume that the model covariance
is identifiable.

We are given a data set consisting of $N$ distinct
individuals, and the measurement vector of the $k$th individual, $\ybt_k$,
has $n_k$ components. $\ybt_k$ is a realization of $\ybf_k$, where
$\ybf_k$ has the Gaussian block linear mixed model structure of Eq.~(1).  
Our problem is to infer $\albf$ and $\theta$ conditional on $\ybt_k$.
The maximum likelihood estimator of $\albf$ for a fixed value of $\theta$ is
\BEQ \label{E3}     
\hat{\albf} = \left( \sum_{\ell=1}^N {\bf X}_{\ell}^t \Sinvl
{\bf X}_{\ell} \right)^{-1} \sum_{k=1}^N 
{\bf X}_k^t \Sinvk \ybt_k 
\ ,
\NEQ     
which has covariance $\Ombf$:
\BEQ \label{E4} 
\Ombf = \left( \sum_{\ell=1}^N {\bf X}_{\ell}^t \Sinvl
{\bf X}_{\ell}
\right)^{-1}
\ . 
\NEQ      
To estimate $\theta$, we maximize the restricted log likelihood functional 
as described in $[1,3-9,12,13,19]$. For block linear mixed models, 
the REML functional is given by
\begin{eqnarray} \label{E5A}     
\ell ( \theta,\albh ) = {C}({N}_T - p) + {1 \over 2}
\sum_{k=1}^N \ln ({\rm det} \ ({\bf X}_k^t {\bf X}_k ))
- \ {1 \over 2} \sum_{k=1}^N \ln ({\rm det} \ ( \Sik ))
\nn \\ 
- {1 \over 2} \ln \left( {\rm det} \ \left( \sum_{k=1}^N {\bf X}_k^t
\Sinvk {\bf X}_k \right) \right)
- \ {1 \over 2} \sum_{k=1}^N (\ybt_k - {\bf X}_k \hat{\albf} )^t
\Sinvk(\ybt_k - {\bf X}_k \hat{\albf} ) 
\ , 
\end{eqnarray} 
where $\hat{\albf}$ is given by Eq.~(\ref{E3}), ${ N}_T \equiv
\displaystyle{\sum_{k=1}^N} n_k$, and ${C}({ N}_T - p) \equiv
- \ {1 \over 2} ({N}_T - p) \ln (2 \pi )$. 

On any compact set, 
the REML estimate of $\theta$ 
exists, but may not be unique. Our restriction, that 
$\displaystyle{\sum_{k=1}^N} {\bf X}_k^t \Sinvk
{\bf X}_k$ is invertible, implies that  for  any fixed value of $\theta$,
$\albh$ is unique. Kackar and Harville have 
proven that any minimum of the REML estimator is an unbiased estimator
of $\albf$ and $\btbf_k$ \cite{KH81}.

Maximizing $\ell(\theta)$ in Eq.~(\ref{E5A}) is often an expensive and ill-conditioned
problem. In \cite{Riedel90}, 
each individual experiment has hundreds of
observations ($n_k \sim 150$). Using the standard formulation of Eq.~(\ref{E5A}),
{ a single descent step requires $O(\sum_k n_k^3)$ operations.}  
In this note, { we rewrite Eq.~(\ref{E5A}) in a computationally
convenient form which requires only $O(Np^3)$ operations per  step.}

To simplify the restricted ML functional, we assume that the column
space of ${\bf Z}_k$ is contained in the column space of ${\bf X}_k$:
${\bf M}({\bf Z}_k ) \subset {\bf M} ({\bf X}_k )$,
where ${\bf M}$ denotes the column space. This requirement
implies that there are $p \times q$ matrices, ${\bf A}_k$, such that
${\bf Z}_k = {\bf X}_k {\bf A}_k$. We call this subclass of 
block linear mixed models, ``random coefficient models''.
ANOVA models, random constant models with
fixed slopes models and the full random coefficient model all satisfy 
this requirement.

We denote the Moore-Penrose generalized
inverse by $^-$ and denote the projection onto the column space of
a particular matrix, $\Cuu$, by $\Puu(\Cuu)$: $\Puu(\Cuu)\equiv
\Cuu(\Cuu^t \Cuu)^- \Cuu^t$,  and define $\Puu_k\equiv\Puu(\Xuu_k)$
When ${\bf M}({\bf Z}_k ) \subset {\bf M} ({\bf X}_k )$, 
the estimate of $\Puu_k\albf$ from the $k$th individual
simplifies to the ordinary least squares estimator:
\BEQ \label{E6}     
\hat{\albf}_k  \equiv 
( {\bf X}_k^t \Sinvk {\bf X}_k)^{-} {\bf X}^t
\Sinvk \ybt_k = ({\bf X}_k^t {\bf X}_k )^- {\bf X}_k^t \ybf_k 
\ . 
\NEQ      
When $\Puu_k$ is not of full rank, $\hat{\albf}_k$ estimates
only ${\Puu_k\albf}$.  
We define the following matrices:
${\bf E}_k \equiv ({\bf X}_k^t {\bf X}_k)^-$,
${\bf F}_k \equiv ({\bf Z}_k^t {\bf Z}_k )^-$, 
$\Kuu_k \equiv \Euu_k \Xuu_k^t \Zuu_k$, and
$\Luu_k \equiv \Fuu_k \Zuu_k^t \Xuu_k \ .$
When $\Zuu_k = \Xuu_k$, $\Kuu_k$ and $\Luu_k$ are
the $p \times p$ projection matrix, $\Puu(\Xuu_k^t\Xuu_k)$. 
We define the matrix
${\bf D}_k$ as the projection of ${\bf D} ( \theta )$ onto the
column space of ${\bf Z}_k^t {\bf Z}_k$:
$ \ \ {\bf D}_k ( \theta )  = {\bf P} ({\bf Z}_k^t {\bf Z}_k )
{\bf D} ( \theta ) {\bf P} ( {\bf Z}_k^t {\bf Z}_k )$.
The covariance of the single individual estimate of Eq.~(\ref{E6}) is
\BEQ \label{E7}
{\rm Cov}[ \hat{\albf}_k \hat{\albf}_k^t] = ({\bf X}_k^t
\Sinvk {\bf X}_k )^{-} =
\sigma_k^2 {\bf E}_k + \Kuu_k \Duu_k \Kuu_k^t
\ ,
\NEQ     
where we have used the Sherman-Morrison-Woodbury identity.
We define the matrix, $\Muu_k$, to be 
\BEQ \label{E8} 
\Muu_k \equiv {\bf X}_k^t\Sinvk {\bf X}_k  =
\sigma_k^{-2} \left( \Xuu_k^t{\bf X}_k - \Luu_k^t \Zuu_k^t\Zuu_k \Luu_k \right)
+ \Luu_k^t[ \sigma_k^2\Fuu_k +\Duu_k ]^{-} \Luu_k
, \NEQ   
In deriving Eq.~(\ref{E8}), we use
the matrix identity:
\BEQ \label{E9}  
( \sigma_k^2 \Iuu_{n_k} + {\bf Z}_k {\bf DZ}_k )^{-1} = \sigma_k^{-2}
[{\bf I}_{n_k} -{\bf Z}_k {\bf F}_k {\bf Z}_k^t ] + {\bf Z}_k {\bf F}_k
[ \sigma_k^2 {\bf F}_k + {\bf D}_k ]^{-} {\bf F}_k {\bf Z}_k^t
\ .   \NEQ   
An alternative representation of $\Muu_k$ can be derived by
applying the Sherman-Morrison matrix identity \cite[p.~33]{Rao71} 
to Eq.~(\ref{E7}).  
When $\Zuu_k = \Xuu_k$, $\Muu_k$ simplifies 
considerably to $\Muu_k=
[\sigma_k^2 {\bf F}_k + \Duu_k]^- =[\sigma_k^2 {\bf E}_k + \Duu_k]^-$.

The ML estimate of $\albh$, Eq.~(3), may be expressed as the weighted
sum of the $N$ individual estimates, $\albh_k$:
\BEQ \label{E10}      
\hat{\albf} = \left( \sum_{\ell=1}^N ({\bf X}_{\ell}^t
\Sinvl {\bf X}_{\ell}) \right)^{-1} 
\sum_{k=1}^N {\bf X}_k^t \Sinvk
\ybt_k =\Omuu \sum_{k=1}^N \Muu_k \albh_k  
\ , \NEQ    
\ni 
where $\Omuu$, the covariance  matrix of $\albh$, satisfies
$\Omuu =  \left( \sum_{k=1}^N {\Muu_k}  \right) ^{-1} $ .

\bigskip 
\ni
{\bf Theorem:} When ${\bf M}({\bf Z}_k ) \subset {\bf M} ({\bf X}_k )$,
the REML functional of Eq.~(\ref{E5A}) reduces to
\begin{eqnarray} \label{E11}
\ell( \theta,\hat{\albf} ) = {C}({N}_T - p) + {1 \over 2} \sum_{k=1}^N
\ln ( {\rm det} \ ({\bf X}_k^t {\bf X}_k ))
- {1\over 2}\sum_{k=1}^N (n_k -q) \ln \sigma_k^2
\nn \\ 
 - {1 \over 2}\sum_{k=1}^N
\ln ( {\rm det} \ ( \sigma_k^2 {\bf I}_q + {\bf Z}_k^t
{\bf Z}_k {\bf D} ))
- {1 \over 2}\ln \left( {\rm det} \left( \sum_{k=1}^N \Muu_k\right) \right)
\nn \\ 
- \sum_{k=1}^N {(n_k - p_k ) \over 2}
{\hat{\sigma}_k^2 \over \sigma_k^2}
- {1 \over 2}\sum_{k=1}^N ( \hat{\albf}_k - \hat{\albf} )^t \Muu_k
( \hat{\albf}_k - \hat{\albf} ) \ ,
\end{eqnarray} 
where
$$(n_k - p_k ) \hat{\sigma}_k^2 \equiv
\ybt_k^t ({\bf I}_k - {\bf P}_k)\ybt_k \ .$$
\medskip

{\it Proof:} We divide the residuals,
$\ybt_k - {\bf X}_k \hat{\albf} = \ybt_k - {\bf X}_k \hat{\albf}_k
+ {\bf X}_k ( \hat{\albf}_k - \hat{\albf} )$, into two parts,
Since $\ybt_k - {\bf X}_k \hat{\albf}_k =({\bf I}_{n_k} - {\bf P}_k)\ybt_k$
is perpendicular to ${\bf X}_k$, the two parts are independent.  
From Eq.~(\ref{E9}), we have
\BEQ \label{E12} 
(\ybt_k - {\bf X}_k \hat{\albf} )^t \Sinvk (\ybt_k - {\bf X}_k
\hat{\albf} ) = (n_k -p_k) \ {\hat{\sigma}_k^2 \over
\sigma_k^2} + ( \hat{\albf}_k - \hat{\albf} )
{\bf M}_k ( \hat{\albf}_k - \hat{\albf} ) 
\ . \NEQ    
A matrix determinant identity yields
\BEQ \label{E13}
\ln ({\rm det} \ ( \Sik )) = (n_k -q) \ln \sigma_k^2 +
\ln ({\rm det} \ ( \sigma_k^2 {\bf I}_q + {\bf Z}_k^t {\bf Z}_k
{\bf D} )) 
\ .
\NEQ 
\  $\Box$

Thus we have reduced the likelihood function from a function of ${N}$
matrices of dimension $n_k$ to ${N}$ matrices of dimension $p$. 
When the individual variances, $\sigma_k^2$, are given, the
likelihood has a simple interpretation as the restricted likelihood of a
set of independent regression coefficients with a normal distribution,
$\hat{\albf}_k \sim {\bf N} ( \Puu_k{\albf} , \sigma_k^2 {\bf E}_k
+ \Kuu_k{\bf D}_k\Kuu_k^t )$.

\medskip \ni

{\bf II. Scoring Algorithm}

Differentiating the REML function with respect to the
$i$th component of $\th$ yields:

\BEQ \label{E14} 
{\partial \ell \over \partial \th_i}={1\over 2}Trace\left[
\sum_{k=1}^N \left( \Guu_k
\left((\albh_k - {\albh} )
(\albh_k - {\albh} )^t +\Omuu \right)
\Guu_k^t -
( \sigma_{k}^2 \Fuu_k + \Duu_k )^{-}  \right) 
{\partial \Duu_k\over \partial\th_i} \right] 
,\NEQ  
where $\Guu_k$ is the $q \times p$ matrix, 
$\Guu_k \equiv (\sigma_k^2 \Fuu_k + \Duu_k )^{-}\Fuu_k\Zuu_k^t \Xuu_k $.
When $\Zuu_k = \Xuu_k$, $\Guu_k$ simplifies 
considerably to $\Guu_k=
[\sigma_k^2 {\bf F}_k + \Duu_k]^- =[\sigma_k^2 {\bf E}_k + \Duu_k]^-$.


From the representation of $\albh$ as the weighted sum of the 
individual $\al_k$, the expectation of the empirical covariance of
the random coefficients satisfies
\BEQ \label{E15} 
\Luu_k\left( E [ (\albh_k - {\albh} )
(\albh_k - {\albh} )^t] + \Omuu_k\right) \Luu_k^t =
\Duu_k + \sigma^2_k \Fuu_k  \ . 
\NEQ 
Thus the REML estimate of $\Duu(\th)$ is a variance weighted version of
``total variance = within individual variance + between individual 
variance.''

By using Eq.~(\ref{E15}) to
compute $\nabla_{\th}\ell(\th,\albh|\ybt)$ the operations count is reduced
to $O(Np^3)$ per step. Similar  savings are achieved in evaluating
the Hessian of the REML. A popular maximization technique is the
scoring method of Fisher, where 
${\partial^2 \ell(\th,\albh|\ybt) \over \partial \th_i\partial \th_j}$
is  replaced with
${\rm E}\left[{\partial^2 \ell \over \partial \th_i\partial \th_j}\right]$.
Thus the  scoring algorithm is 
$$\th^{new} = \th^{old} + \Juu^{-1} \nabla_{\th}\ell(\th^{old},\albh|\ybt)
,\eqno (16)$$
where $\Juu_{i,j} \equiv -
{\rm E}\left[{\partial^2 \ell \over \partial \th_i\partial \th_j}\right]
(\th^{old},\albh|\ybt)$ with
$${\rm E}\left[{\partial^2 \ell (\th^{old},\albh|\ybt)
\over \partial \th_i\partial \th_j}\right]=
{-1\over 2} \sum_{k=1}^N 
Trace\left[ \Guu_k \left(
{\partial \Duu_k\over \partial\th_i} -
{\partial \Omuu\over \partial\th_i}  \right)
\Guu_k{\partial \Duu_k\over \partial\th_j} \right] 
.\eqno (17)$$
We initialize the scoring algorithm at $\th=0$.
\medskip

Remarks:

1) Swamy \cite{Swamy71} 
has derived the analogous expression for the ML function of the
random coefficient model under the assumptions that $\Zuu_k \equiv \Xuu_k$
and that the ${\bf X}_k^t{\bf X}_k$ are nonsingular.

2)
The within individual estimate of $\sigma_k^2$ is
\BEQ \label{E18}    
\hat{\sigma}_{k}^2  = {\ybt_k^{\ t} \Puu_{\perp k} \ybt_k \over
n_k - m_{k} }, 
\NEQ     
\ni
where 
$\Puu_{\perp k}$ is the projection perpendicular
to the extended column space of $(\Xuu_k, \Zuu_k)$:
$\Puu_{\perp k} \equiv \Iuu_{n_k} - \Puu(\Xuu_k, \Zuu_k)$,
and $m_k$ is the number of degrees of freedom used in the fit,
 i.e. the rank of $\Puu(\Xuu_k, \Zuu_k)$.
This estimate is consistent as $n_k \rightarrow \infty$.

The REML estimates for $\sigma_k^2$ are noticeably more
complicated than the simple within individual least squares estimate
of Eq.~(\ref{E18}). 
Hybrid estimation schemes, which utilize the REML Eqs.~(\ref{E10}) 
and (\ref{E14}) to
estimate $\al$ and $\Duu(\th)$, and Eq.~(\ref{E18}) to estimate $\sigma_k^2$,
are of considerable practical interest. This hybrid estimate
is empirical Bayesian in $\sigma_k^2$.
 
3) Since $\Duu(\th)$ is the variance weighted
difference of the empirically estimated covariance of the $\albh_k- \albh$
matrices, and the ordinary least squares estimate of the within
individual variance,
it can have negative eigenvalues. The standard
solution to this problem is to set the negative eigenvalues to zero
using the singular value decomposition. This procedure corresponds to
imposing a positivity constraint on $\Duu(\th)$. The constraint
produces a slight positive bias in the estimate of $\Duu(\th)$.
The REML estimate of $\Duu(\th)$ is larger than the ML estimate
of $\Duu(\th)$, and thus the REML correction
reduces the probability of negative eigenvalues, but does not eliminate
it. 

4) When 
data are missing, the expected maximization 
(EM) algorithm
may be applied to Eqs.~(\ref{E10}), and (\ref{E14}) directly just as the EM
algorithm has been applied to the original REML formulation $[9,10,12,13,19]$.

5)
The REML estimator does not directly use estimates of the 
random effects, $\btbf_k$, nor does it yield estimates of $\btbf_k$.
However, 
the random effects may be estimated from 
$\bth_k = {\bf D}_k
{\bf Z}_k^t\Sibf_k^{-1} (\ybt_k - {\bf X}_k \hat{\albf} ) = 
{\bf D}_k [ \sigma_k^2\Fuu_k +\Duu_k ]^{-} \Luu_k (\albh_k - \hat{\albf} ),$ 
which is both the best linear unbiased estimator when $\th$ is known and
the empirical Bayesian estimator when $\th$ is estimated from Eq.~(11).

{\bf III. Conclusion}

The restricted maximum likelihood
estimator for the dispersion matrix of random effects models 
requires only the ordinary least squares estimates
for each separate individual and the corresponding covariance
matrices of the individual estimates. Thus we have reduced
the computational cost per descent step from $O(\sum_k n_k^3)$ operations  
to $O(Np^3)$ operations per  step.

{\it Acknowledgment}

This work was supported by the U.S. Department of Energy.


\end{document}